\documentclass[conference]{IEEEtran}

\ifCLASSOPTIONcompsoc
\usepackage[nocompress]{cite}
\else
\usepackage{cite}
\fi

\ifCLASSINFOpdf
\usepackage[pdftex]{graphicx}
\graphicspath{{./image/}}
\DeclareGraphicsExtensions{.pdf,.jpeg,.png}
\else
\usepackage[dvips]{graphicx}
\graphicspath{{./image/}}

\DeclareGraphicsExtensions{.eps}
\fi

\usepackage{algorithm}
\usepackage{graphicx,algorithm,algorithmicx,bm,amsmath,amssymb,rotating,graphicx,cite,amsthm,calc,color,epstopdf,xcolor,mathrsfs,dsfont}
\usepackage{multirow}
\usepackage{graphicx,caption}
\usepackage{algpseudocode}

\usepackage{amsmath}
\usepackage{lettrine}
\usepackage{graphicx,color}
\usepackage{colortbl}
\usepackage{cite}
\usepackage{longtable}
\usepackage{amsmath}
\usepackage{float}
\usepackage{array}
\usepackage{tikz}
\usepackage{verbatim}
\usepackage{amsmath}
\usepackage{algpseudocode}
\usepackage{caption}
\usepackage{subcaption}
\usepackage{pifont}
\usepackage{amssymb}
\usepackage{mathtools}
\usepackage{verbatim}
\usepackage{soul}
\usepackage{makecell}
\usepackage{flushend}
\usepackage{enumitem}

\newcolumntype{P}[1]{>{\centering\arraybackslash}p{#1}}
\newcolumntype{M}[1]{>{\centering\arraybackslash}m{#1}}

{}
\makeatletter

\def\endthebibliography{%
\def\@noitemerr{\@latex@warning{Empty `thebibliography' environment}}%
\endlist
}
\makeatother
\ifCLASSINFOpdf

\else

\fi

\def\F{\bm{F}}
\def\I{\bm{I}}
\def\FM{\text{MSMV-Swin}}

\usepackage[left=0.64in,right=0.64in,top=0.8in, bottom=1.1in]{geometry}

\begin{document}

\author{\IEEEauthorblockN{Farnoush Bayatmakou\IEEEauthorrefmark{1},
Reza Taleei\IEEEauthorrefmark{2},
Milad Amir Toutounchian\IEEEauthorrefmark{3},
Arash Mohammadi\IEEEauthorrefmark{1}}

\IEEEauthorblockA{\IEEEauthorrefmark{1}Concordia Institute for Information Systems Engineering (CIISE), Concordia University, Montreal, Canada}
\IEEEauthorblockA{\IEEEauthorrefmark{2}Thomas Jefferson University Hospital, Philadelphia, Pennsylvania, USA}

        \IEEEauthorblockA{\IEEEauthorrefmark{3}
        College of Computing \& Informatics, Drexel University, Philadelphia, Pennsylvania, USA}
}
\title{Integrating AI for Human-Centric Breast Cancer Diagnostics: A Multi-Scale and Multi-View Swin Transformer Framework}

\IEEEtitleabstractindextext{%
\begin{abstract}
Despite advancements in Computer-Aided Diagnosis (CAD) systems, breast cancer remains one of the leading causes of cancer-related deaths among women worldwide.  Recent Artificial Intelligence (AI) breakthroughs have shown significant promise in developing advanced Deep Learning (DL) architectures for breast cancer diagnosis through mammography. In this context, the paper focuses on integrating AI within a Human-Centric workflow to enhance breast cancer diagnostics. Key challenges are, however, largely overlooked, such as reliance on detailed tumor annotations and susceptibility to missing views, particularly during test time. To address these issues, we propose a hybrid, multi-scale and multi-view Swin Transformer-based framework ($\FM$) that enhances diagnostic robustness and accuracy. The proposed $\FM$ framework is designed to be a decision-support tool, helping radiologists analyze multi-view mammograms more effectively. More specifically, the $\FM$ framework leverages the Segment Anything Model (SAM) to isolate the breast lobe, reducing background noise and enabling comprehensive feature extraction. The multi-scale nature of the proposed $\FM$ framework accounts for tumor-specific regions and the spatial characteristics of tissues surrounding the tumor, capturing both localized and contextual information. Integrating contextual and localized data ensures that $\FM$'s outputs align with how radiologists interpret mammograms, fostering better human-AI interaction and trust. A hybrid fusion structure is then designed to provide robustness against missing views, a common occurrence in clinical practice when only a single mammogram view is available. Experimental evaluations on single-view and dual-view mammography based on CBIS-DDSM dataset demonstrate the superior performance of $\FM$, highlighting its potential for improving breast cancer diagnosis in diverse clinical settings.
\end{abstract}
}

\maketitle

\IEEEdisplaynontitleabstractindextext

\ifCLASSOPTIONcompsoc
\IEEEraisesectionheading{\section{Introduction}\label{sec:introduction}}
\else
\section{Introduction}\label{sec:introduction}
\fi 
Breast cancer is one of the leading causes of cancer-related deaths among women worldwide, making early detection of significant importance~\cite{obeagu2024breast}. Multi-view mammography~\cite{jouirou2019multi} has emerged as a vital approach to breast cancer diagnosis, where radiologists analyze multiple views (e.g., CranioCaudal (CC) and MedioLateral Oblique (MLO)) to detect abnormalities that may not be visible in a single view. This has inspired the development of multi-view-based Computer-Aided Diagnosis (CAD) systems, which utilize information from different views to improve diagnostic accuracy. In recent years, Deep Learning (DL) models, particularly Convolutional Neural Networks (CNNs)~\cite{khan2019multi, sun2019multi} and Transformers~\cite{zhao2020cross, chen2022transformers}, have significantly enhanced the performance of multi-view CAD systems. Despite recent advancements in this domain, however, key challenges remain ahead, including reliance on detailed annotated Region of Interest (ROIs) and susceptibility to missing views, especially at the test time. Addressing these limitations is critical to advancing CAD system reliability and robustness. This paper proposes a hybrid, multi-view, and multi-scale framework that leverages state-of-the-art DL techniques to overcome these challenges. The proposed approach enhances diagnostic precision and robustness, providing a reliable tool for breast cancer diagnosis in diverse clinical contexts.

\subsection{Related Works}
CNN-based approaches dominated initial efforts in multi-view mammogram classification. For instance, Reference~\cite{carneiro2017deep} was among the first to apply multi-view analysis by training CNN models separately on CC and MLO views and combining features using multinomial logistic regression. Such initial works demonstrated that multi-view analysis has the potential to outperform single-view models by leveraging the inherent correlations between different views. Subsequent research improved on these ideas by developing more effective fusion strategies. For example, Reference~\cite{wu2019deep} introduced a view-wise feature merging strategy using tailored ResNet models to process CC and MLO images separately before averaging their predictions during inference. In another notable effort, Reference~\cite{khan2019multi} proposed a two-stage multi-view fusion strategy for extracting ROI from four mammogram views using CNNs, further enhancing the classification performance by integrating multi-view information.

More recent developments have shifted towards using Transformers for multi-view mammography due to their superior ability to model long-range dependencies and capture inter-view correlations. For instance, Reference~\cite{van2021multi} introduced a hybrid approach that integrated CNN feature extraction with global cross-view attention via Transformers. The cross-view fusion focused on correlating CC and MLO views to improve diagnostic accuracy, although it did not fully explore bilateral asymmetry. In another noteworthy study, Reference~\cite{chen2022transformers} used a pure Vision Transformer (ViT) architecture for multi-view breast cancer detection, which showed promising results. This model processed views later in the network, focusing on retaining local features while integrating global multi-view information; however, it still missed some early correlations essential for optimal classification. The introduction of the Swin Transformer~\cite{liu2021swin} architecture, specifically designed for computer vision tasks, has been transformative in multi-view breast cancer classification. Recently, Reference~\cite{sarker2024mv} proposed a pure Transformer-based multi-view architecture built upon the Swin Transformer, incorporating novel shifted window attention mechanisms to integrate multi-views at the spatial feature level. The model captured critical cross-view correlations by fusing information from the CC and MLO views early in the network, significantly outperforming traditional CNN and hybrid models. 

Despite the above-mentioned surge of interest in DL-based multi-view breast cancer detection and achieving exceptional results, the following two significant challenges have been overlooked in the literature, which are the focus of this study: 
\begin{itemize}[noitemsep, nolistsep]
\item[(i)] Most existing models work best when provided with cropped mammograms focused on the tumour region, which ignores the spatial properties of the tissues surrounding the tumor. Such properties are crucial for radiologists and should be incorporated into a DL pipeline.
\item[(ii)] While multi-view DL architectures showed superiority against their single-view counterparts, it is common to have cases with only a single view. Robustness against missing views at test time has been mostly overlooked in the literature. 
\end{itemize}
The paper targets addressing these issues as outlined below.

\setlength{\textfloatsep}{0pt}
\begin{figure}
\centering
\includegraphics[width=0.4\linewidth]{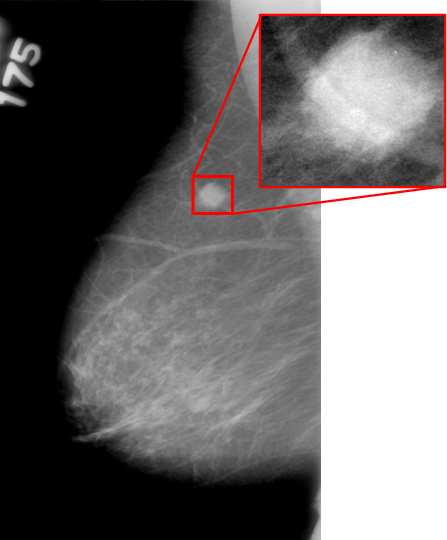}
\caption{\footnotesize Sample image with the cropped area around the designated ROI}
\label{fig: sample_img}
\end{figure}

\noindent

\subsection{Contribution}
Capitalizing on the above discussion, the paper introduces a hybrid, multi-scale, and multi-view Swin Transformer-based architecture called the $\FM$ framework. The multi-scale nature of the $\FM$ framework is introduced following the success of our multi-scale learning architecture, the 3D-MCN~\cite{afshar20203d} in lung nodule malignancy. The multi-scale learning is incorporated to consider the spatial characteristics of tissues surrounding the tumor, addressing the first identified issue. The $\FM$ framework is further designed to tackle the second challenge by introducing robustness against missing views during test time through a feature zero padding strategy. More specifically, the proposed framework incorporates different scales of each mammogram. The segmented scale is obtained using the Segment Anything Model (SAM)~\cite{kirillov2023segment}, capable of segmenting any discernible feature within a mammogram. This capability is crucial for extracting detailed representations of breast tissues in both CC and MLO views. As shown in Fig.~\ref{fig: sample_img}, two scales are considered: (i) \textit{Masked scale}, which is obtained by segmented bounding boxes around the identified breast lobe, and (ii) \textit{Cropped scale}, which focuses on regions around the pre-existing ROI with tumors. Cropped images provide enhanced details and can improve accuracy in analyzing critical tumor regions. Conversely, using a segmented scale allows for an extensive examination of both the localized tumor areas and the broader anatomical features, including their position, shape, and relative size within the breast anatomy. Such a multi-scale radiomics approach provides essential insights into the interactions between tumors and their surrounding tissues, critical for delivering precise diagnostic outcomes. 
To optimize feature extraction from both CC and MLO views, we employ a pre-trained model, Shifted Windows (Swin) Transformer~\cite{liu2021swin}. These models facilitate robust feature extraction, which is further enhanced through various feature fusion techniques. The fused features are then channeled through a sequence of Multi-Layer Perceptrons (MLPs) to achieve final classification results.


\section{Problem Formulation and Proposed  $\FM$ Framework} \label{sec:FM}
\begin{figure*}[t!]
\centering
\includegraphics[scale=.54]
{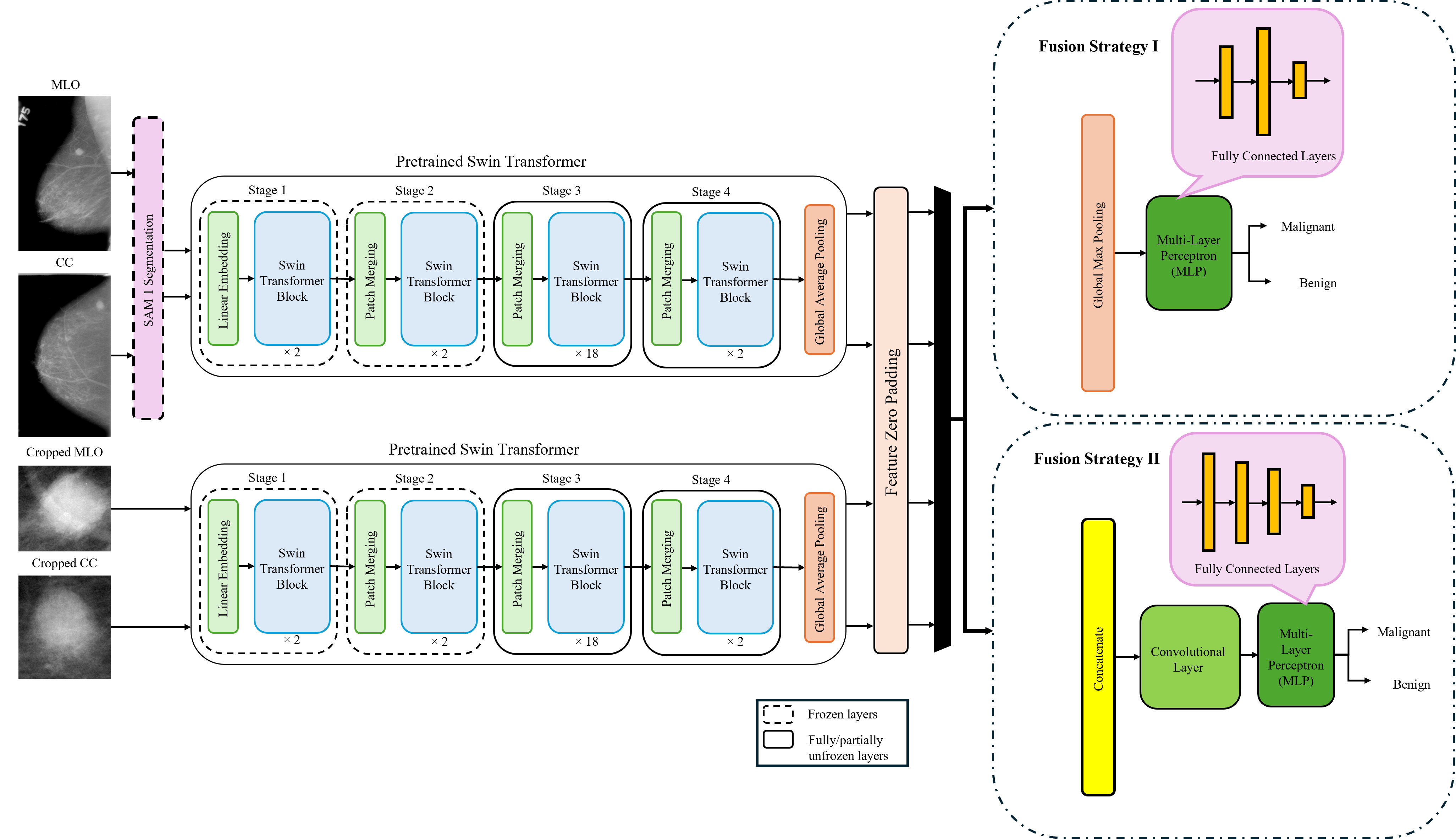}
\caption{\footnotesize The proposed $\FM$ framework with feature fusion through (I) max-pooling-based fusion and (II) convolutional and MLP-based fusion.}
\label{fig:swin-conv, swin-maxpool}
\end{figure*}
The $\FM$ framework leverages a dual Swin Transformer architecture to process distinct scales of mammographic image inputs—specifically, cropped and segmented variants. The processing pipeline is designed to handle separate streams for CC and MLO views for each of the two underlying scales. This dual-path approach ensures comprehensive feature extraction tailored to the specific characteristics of each image type.

\subsection{Model Architecture and Training}
As the pre-processing step, we utilized a well-trained SAM to segment original mammograms. This approach is beneficial in enhancing the learning process of the $\FM$ framework by removing the background areas from mammogram images to isolate the breast lobe.

The extracted breast lobe is then passed into two parallel paths of the proposed $\FM$ framework. More specifically, extracted breast areas after pre-processing are down-sampled to ($224, 224$) pixels to reduce the complexity and memory allocation without significant loss of information. Afterwards, we leverage a pre-trained Swin Transformer for feature extraction, which is known for its efficacy in processing complex image structures. We have also used two different feature fusion strategies to combine the extracted features to be fed to an MLP for the final classification step.

The feature extraction and processing pipeline is divided into two distinct paths: (i) \textit{Segmented Image Path}: This branch processes features from segmented CC and MLO views,
forming a comprehensive feature set that provides a complete representation of each segmented image, facilitating a detailed analysis across various diagnostic image types, 
and (ii) \textit{Cropped Image Path}: Similarly, this pathway focuses on features extracted from specifically cropped areas within the CC and MLO views.
Considering these features ensures that critical areas of interest are emphasized, enhancing the scrutiny and diagnostic potential for regions that may exhibit pathological features.

\subsection{Feature Extraction Framework}
Pre-trained Swin Transformer is deployed in dual configurations given its capabilities to process images hierarchically and efficiently capture long-range dependencies through a shifted window strategy, which is essential for integrating local and global contextual features effectively. The architecture integrates dual instances of Swin Transformers, each fit to the unique characteristics of the corresponding scale and set to process segmented and cropped images independently. Such an approach facilitates precise feature extraction from essential regions, making it particularly suitable for detailed medical diagnostics of the two views. Initially, all parameters in each Swin Transformer instance are frozen, except the top layers, which are selectively unfrozen during fine-tuning to preserve learned general features while adapting to the underlying dataset's specific characteristics. Strategic unfreezing of a subset of the last layers during the training phase customizes the model to the special features of CC and MLO images and the targeted areas in cropped images.

After feature extraction using Swin Transformers, each of the two instances associated with the two scales processes two views, resulting in four feature sets of dimension $1,024$. Let these features be denoted by: (i) $\F^{Seg}_{\text{CC}}$: Features of segmented CC images; (ii) $\F^{Seg}_{\text{MLO}}$: Features of segmented MLO images; (iii) $\F^{Crop}_{\text{CC}}$: Features of cropped CC images, and;  (iv) $\F^{Crop}_{\text{MLO}}$: Features of cropped MLO images. Moreover, we applied a feature zero padding approach before the feature fusion step to handle cases with missing views. The corresponding feature vector is set to zero in scenarios where either the CC or MLO view is unavailable. This ensures that no invalid data is propagated through the network while allowing the available view to contribute fully to the final feature representation. By preserving the integrity of the feature fusion process, this approach maintains the framework's robustness even when dealing with incomplete input data, a common occurrence in clinical practice. The final outputs of the feature extraction framework are given by
\begin{equation}
\label{eq: fzeropad}
    \F^{Seg}_{\text{CC}}=\begin{cases}
        \text{Swin}\left(\I^{Seg}_{\text{CC}}\right) & \text{if}~\text{CC view exists}\\
        0, & \text{otherwise}
    \end{cases}
\end{equation}
where $\I^{Seg}_{\text{CC}}$ represents the segmented mammogram image from the CC view. The other feature vectors, after applying feature zero padding, are defined in a similar manner.

\subsection{Fusion Framework}
 After extracting the features from the existing views, the four feature vectors are subsequently fused to form a unified feature map for downstream processing, given by
\begin{eqnarray}
\F_{\text{combined}} = f \left( \F^{Seg}_{\text{CC}}, \F^{Seg}_{\text{MLO}}, \F^{Crop}_{\text{CC}}, \F^{Crop}_{\text{MLO}} \right),
\end{eqnarray}
where function $f(\cdot)$ is the fusion function.
The following two fusion strategies are implemented.

\subsubsection{Max-Pooling Fusion}
Under this fusion strategy, the feature vector $\F_{\text{combined}}$ is constructed by concatenating the four features extracted from cropped and segmented mammograms across CC and MLO views. The resultant feature map undergoes a reduction process where the maximum values are selected across the concatenated features. This operation is performed across dimensions with different views (CC and MLO) and segmentation statuses (segmented and cropped), effectively capturing the most critical features. This max-pooling step ensures that only the most relevant features with the highest values post concatenation are passed forward, enhancing the model's focus on potentially critical diagnostic information.

The reduced feature map is then directed through a robust MLP classification head, which comprises successive layers of linear transformations and non-linear activation functions. Specifically,  LeakyReLU is used to facilitate smooth gradient flow together with dropout layers to mitigate overfitting. The architecture of the MLP features layer sizes that progressively decrease from 1,024 to 512. The idea is to ensure the refinement of the feature set into highly discriminative elements essential for accurate classification.
%
Following the feature compression down to 512 units, the model's architecture culminates in a single output layer. This layer directly projects the 512-dimensional feature vector to a single logit, indicative of the binary classification task.

\subsubsection{Convolution-based Fusion}
In this fusion strategy, the outputs of the dual transformers are concatenated into a $1024\times 4$ array and passed through a custom-designed convolutional block. A convolutional layer further refines extracted features by enhancing spatial relationships and reducing dimensionality, thus focusing the model's attention on the most significant aspects of the image. This combination not only maximizes the feature extraction capabilities of Swin Transformers but also leverages the robust, spatial feature processing of convolutional networks, significantly improving the accuracy and reliability of image classification tasks by ensuring a comprehensive and refined analysis of input data. The incorporated block consists of a 2D convolutional layer with a kernel size of $3 \times 3$, followed by batch normalization, a ReLu activation function, and a max-pooling layer.

As previously described, the refined features are flattened and fed into an MLP layer.
It is to be noted that, as the only difference, the input from the convolution block to the MLP is of size $4096$.
\section{Experimental Result}\label{sec:results}
\subsection{Dataset}
The proposed $\FM$ framework is developed based on the Curated Breast Imaging Subset of the Digital Database for Screening Mammography (CBIS-DDSM) dataset~\cite{CBIS-DDSM:2017}. CBIS-DDSM is a widely recognized mammography dataset and comprises $10,239$ images, including whole mammograms, cropped images, and ROI mask images with mass and calcification. A sample of mammograms from the CBIS-DDSM and its cropped version in the designated ROI is illustrated in Fig.~\ref{fig: sample_img}. To achieve results comparable to other studies utilizing the CBIS-DDSM dataset, we assigned patients to train and test subsets following the splitting presented in~\cite{CBIS-DDSM:2017}. We extracted cropped and full mammogram images from the CBIS-DDSM dataset, selecting patients with distinct CC and MLO views. The corresponding CC and MLO images were treated as a single examination per breast, resulting in $477$ samples for training and $144$ for testing.


While only samples with both views are used for training, excluding $208$ single-view breast samples from the training set, $59$ single-view cases are included during testing. In other words, although the $\FM$ framework is trained in a multi-view setting, it is also designed to support classification of single-view cases at test time.
Data augmentation was used to increase the training dataset, where each mammogram is augmented by $90$, $180$, and $270$-degree rotations, as well as horizontal and vertical flipping. Following this augmentation approach, five new data points are generated and added to the original breast sample. The same type of augmentation is applied on the CC and MLO views, resulting in an increase in the size of the training and test datasets by a factor of 6 ($477\times6$, $144\times6$, and $59\times6$, respectively, for multi-view training and test datasets, and single-view test dataset).

To compensate for the missing views (missing modality) at the test-time (patients with either CC or MLO view), different approaches can be used, borrowing from the works on missing modality in multi-modal learning~\cite{wang2023multi, MM-Imputation:2023, ma2021maximum, shi2024incomplete}. One naive approach is to zero-pad the missing view. This, however, is expected to deteriorate the results as observed through our empirical evaluations. An alternative approach could be to simulate the missing view based on the available one. In other words, the missing data can be imputed using generative models; the imputation process, however, may introduce unrealistic information to the classification process, leading to poor performance. Instead, we opted for feature zero padding for the missing view. This method ensures that the representation remains consistent and avoids introducing potentially misleading synthetic data. Furthermore, by focusing on the feature space rather than the raw input space, we mitigate the negative impact of missing modalities while preserving the integrity of the classification process.


\subsection{Results}

The model is trained using a binary cross-entropy with a label smoothing loss function. This loss function helps improve model generalization by penalizing the model less severely on complex examples and smoothing the decision boundary. We used AdamW optimizer, chosen for its ability to combine the benefits of Adam optimization and weight decay regularization, making it particularly suited for this kind of complex model training. Validation is conducted in parallel with training using a separate subset of the data. This approach allows continuous monitoring of the model's performance and generalization capabilities, ensuring robustness before deployment.
To evaluate model performance across varying thresholds and conditions comprehensively, we report standard classification metrics, including accuracy, sensitivity, specificity, and the area under the ROC curve (AUC).
Table~\ref{Table1} illustrates the results obtained from Swin-based back-ends. These results compare two fusion mechanisms and the model's performance with missing views. As can be observed, max-pooling fusion seems to provide slightly better overall performance across several metrics. This is an interesting observation, showing that complex fusion is not required, which can contribute to the strength of the feature extraction pipeline. At the same time, it is observed that the performance degradation resulting from missing views is relatively limited, illustrating the robustness of the proposed pipeline. Finally, Table~\ref{table:tsota} illustrates the comparison results with recent state-of-the-art~\cite{sarker2024mv, dos2024deep, quintana2023exploiting}, further corroborating the superiority of the proposed $\FM$ framework. 


\begin{table}[t!]
\centering
\caption{$\FM$ framework with Swin-based back-end. \label{Table1}}
\resizebox{\columnwidth}{!}{
\begin{tabular}{|l|c|c|c|c|c|}
\hline
\textbf{Test Data} & \textbf{Accuracy (\%)} & \textbf{AUC (\%)} & \textbf{F1 Score (\%)} & \textbf{Sensitivity (\%)} & \textbf{Specificity (\%)} \\ \hline
\multicolumn{6}{|c|}{\textbf{Swin with Max-Pooling Fusion}}
 \\ \hline
Breasts with both Views & \textbf{80.32} & 82.27 & 75.07 & 72.32 & \textbf{85.88} \\ \hline
Breasts with Missing Views & 77.40 & \textbf{84.20} & 74.36 & \textbf{80.56} & 75.24 \\ \hline
All Breasts         & \textit{79.06} & \textit{81.50} & \textit{74.42} & \textit{74.50} & \textit{82.22} \\ \hline
\multicolumn{6}{|c|}{\textbf{Swin with Convolution-based Fusion}}
 \\ \hline
Breasts with both Views & \textbf{79.05} & 84.67 & 74.54 & 74.86 & \textbf{81.96} \\ \hline
Breasts with Missing Views & 77.68 & \textbf{85.57} & 74.76 & \textbf{81.25} & 75.24 \\ \hline
All Breasts       & 77.50 & 83.73 & 73.19 & 75.10 & 79.17 \\ \hline

\end{tabular}}
\end{table}


\begin{table}[t!]
\caption{\footnotesize Comparison with state-of-the-art.}
\centering
\resizebox{1\columnwidth}{!}{
\begin{tabular}{llccccc}
\Xhline{1.5pt}
\multirow{2}{*}{ Model}                                                                                                                                                                       & \multirow{2}{*}{Variants} & \multirow{2}{*}{Accuracy (\%)} & \multirow{2}{*}{AUC (\%)} & \multirow{2}{*}{F1 Score (\%)} & \multirow{2}{*}{ Sensitivity (\%)} & \multirow{2}{*}{ Specificity (\%)} \\
&                                               &                        &                                                         &                                                             &                                                             &                                                             \\ \hline                                              \\
\rule{0pt}{2.5ex} \multirow{3}{*}{{\begin{tabular}[c]{@{}l@{}}  {$\FM$ Framework}\\  {Max-Pooling Fusion}\end{tabular}}} 
&  Both-Views      & $80.32$ & $82.27$ & $75.07$ & $72.32$ & $85.88$ \\
&  Missing-Data    & $77.40$ & $84.20$ & $74.36$ & $80.56$ & $75.24$ \\
&  All Breasts     & $79.06$ & $81.50$ & $74.42$ & $74.50$ & $82.22$ \\ \hline                                              
\rule{0pt}{2.5ex} \multirow{3}{*}{{\begin{tabular}[c]{@{}l@{}}  {Reference~\cite{sarker2024mv}}\\  {}\end{tabular}}}  &  Both-Views                  & $68.63$                & $71.37$                                                 & -                                                    & -                                                    & -                                                     \\
&  Missing-Data                    & -                & -                                                 & -                                                     & -                                                     & -                                                     \\ \hline
\rule{0pt}{2.5ex} \multirow{3}{*}{{\begin{tabular}[c]{@{}l@{}}  {Reference~\cite{quintana2023exploiting}}\\  {}\end{tabular}}}  &  Both-Views                  & $73.00 \pm 1.90$ &     $80.90 \pm 0.50$                                             &$-$                                                     &$-$                                                     & $71.00 \pm 2.00$                                                     \\
&  Missing-Data                    & -                & -                                                 & -                                                     & -                                                     & -                                                     \\ \hline
\rule{0pt}{2.5ex} \multirow{3}{*}{{\begin{tabular}[c]{@{}l@{}}  {Reference~\cite{dos2024deep}}\\  {}\end{tabular}}}  &  Both-Views                  & $72.7 \pm 1.91$                & $-$                                                 & $-$                                                     & $-$                                                     & $-$                                                     \\
&  Missing-Data                    & -                & -                                                 & -                                                     & -                                                     & -                                                     \\ \hline

\rule{0pt}{2.5ex} \multirow{3}{*}{{\begin{tabular}[c]{@{}l@{}} {Reference~\cite{yang2024mammo}}\\  {}\end{tabular}}}  &  Both-Views                  & $70.9$               & $80.5 \pm 0.2$                                                 & $70.9$                                                     & $-$                                                     & $-$                                                     \\
&  Missing-Data                    & -                & -                                                 & -                                                     & -                                                     & -                                                     \\ \hline
\rule{0pt}{2.5ex} \multirow{3}{*}{{\begin{tabular}[c]{@{}l@{}} {Reference~\cite{allaoui2024hybridmammonet}}\\  {}\end{tabular}}}  &  Both-Views                  & $-$               & $80.0$                                                 & $65.0$                                                     & $-$                                                     & $-$                                                     \\
&  Missing-Data                    & -                & -                                                 & -                                                     & -                                                     & -                                                     \\ \hline
\rule{0pt}{2.5ex} \multirow{3}{*}{{\begin{tabular}[c]{@{}l@{}} {Reference~\cite{liao2024open}}\\  {}\end{tabular}}}  &  Both-Views                  & $68.24$               & $74.21$                                                 & $65.38$                                                     & $62.96$                                                     & $-$                                                     \\
&  Missing-Data                    & -                & -                                                 & -                                                     & -                                                     & -                                                     \\ \hline

\end{tabular}
}
\label{table:tsota}
\end{table}

\section{Conclusion}\label{sec:conc}
In this paper, we introduced the $\FM$ framework, a hybrid, multi-scale and multi-view Swin Transformer-based architecture aimed at addressing two key challenges in breast cancer diagnosis through mammography, i.e., (i) Reliance on detailed tumor annotations and (ii) Susceptibility to missing views during test time. The $\FM$ framework distinguishes itself by utilizing a novel approach that integrates localized and contextual characteristics of tissues surrounding tumors. In other words, the framework processes multi-scale information from both CC and MLO views, effectively capturing both localized and contextual features. Additionally, two feature fusion strategies were employed to ensure robustness by compensating for missing views without relying on synthetic data. Experimental results demonstrate that the $\FM$ framework outperforms several existing models. Notably, it achieves superior accuracy and robustness even in scenarios where only single-view mammograms are available, with only a modest reduction in accuracy compared to dual-view cases. These findings underscore the potential of the $\FM$ framework to advance reliable and comprehensive breast cancer diagnostics in real-world clinical settings.

\ifCLASSOPTIONcaptionsoff
\newpage
\fi

\bibliographystyle{IEEEtran}
\bibliography{refs.bib}

@inproceedings{kirillov2023segment,
  title={Segment anything},
  author={Kirillov, Alexander and Mintun, Eric and Ravi, Nikhila and Mao, Hanzi and Rolland, Chloe and Gustafson, Laura and Xiao, Tete and Whitehead, Spencer and Berg, Alexander C and Lo, Wan-Yen and others},
  booktitle={Proceedings of the IEEE/CVF International Conference on Computer Vision},
  pages={4015--4026},
  year={2023}
}

@article{afshar20203d,
  title={3D-MCN: a 3D multi-scale capsule network for lung nodule malignancy prediction},
  author={Afshar, Parnian and Oikonomou, Anastasia and Naderkhani, Farnoosh and Tyrrell, Pascal N and Plataniotis, Konstantinos N and Farahani, Keyvan and Mohammadi, Arash},
  journal={Scientific reports},
  volume={10},
  number={1},
  pages={7948},
  year={2020},
  publisher={Nature Publishing Group UK London}
}

@article{CBIS-DDSM:2017,
  title={A curated mammography data set for use in computer-aided detection and diagnosis research},
  author={Lee, Rebecca Sawyer and Gimenez, Francisco and Hoogi, Assaf and Miyake, Kanae Kawai and Gorovoy, Mia and Rubin, Daniel L},
  journal={Scientific data},
  volume={4},
  number={1},
  pages={1--9},
  year={2017},
  publisher={Nature Publishing Group}
}

@article{MM-Imputation:2023,
  author={Chen, Yuanyuan and Pan, Yongsheng and Xia, Yong and Yuan, Yixuan},
  journal={IEEE Transactions on Medical Imaging}, 
  title={Disentangle First, Then Distill: A Unified Framework for Missing Modality Imputation and Alzheimer’s Disease Diagnosis}, 
  year={2023},
  volume={42},
  number={12},
  pages={3566-3578}
}

@inproceedings{wang2023multi,
  title={Multi-modal learning with missing modality via shared-specific feature modelling},
  author={Wang, Hu and Chen, Yuanhong and Ma, Congbo and Avery, Jodie and Hull, Louise and Carneiro, Gustavo},
  booktitle={Proceedings of the IEEE/CVF Conference on Computer Vision and Pattern Recognition},
  pages={15878--15887},
  year={2023}
}

@article{ma2021maximum,
  title={Maximum likelihood estimation for multimodal learning with missing modality},
  author={Ma, Fei and Xu, Xiangxiang and Huang, Shao-Lun and Zhang, Lin},
  journal={arXiv preprint arXiv:2108.10513},
  year={2021}
}

@article{shi2024incomplete,
  title={Incomplete Cross-Modal Retrieval with Deep Correlation Transfer},
  author={Shi, Dan and Zhu, Lei and Li, Jingjing and Dong, Guohua and Zhang, Huaxiang},
  journal={ACM Transactions on Multimedia Computing, Communications and Applications},
  volume={20},
  number={5},
  pages={1--21},
  year={2024},
  publisher={ACM New York, NY}
}

@article{sarker2024mv,
  title={MV-Swin-T: Mammogram Classification with Multi-view Swin Transformer},
  author={Sarker, Sushmita and Sarker, Prithul and Bebis, George and Tavakkoli, Alireza},
  journal={arXiv preprint arXiv:2402.16298},
  year={2024}
}

@article{carneiro2017deep,
  title={Deep learning models for classifying mammogram exams containing unregistered multi-view images and segmentation maps of lesions},
  author={Carneiro, Gustavo and Nascimento, Jacinto and Bradley, Andrew P},
  journal={Deep learning for medical image analysis},
  pages={321--339},
  year={2017}
}

@article{wu2019deep,
  title={Deep neural networks improve radiologists’ performance in breast cancer screening},
  author={Wu, Nan and Phang, Jason and Park, Jungkyu and Shen, Yiqiu and Huang, Zhe and Zorin, Masha and Jastrzbski, Stanisaw and Fvry, Thibault and Katsnelson, Joe and Kim, Eric and others},
  journal={IEEE transactions on medical imaging},
  volume={39},
  number={4},
  pages={1184--1194},
  year={2019},
  publisher={IEEE}
}

@article{khan2019multi,
  title={Multi-view feature fusion based four views model for mammogram classification using convolutional neural network},
  author={Khan, Hasan Nasir and Shahid, Ahmad Raza and Raza, Basit and Dar, Amir Hanif and Alquhayz, Hani},
  journal={IEEE Access},
  volume={7},
  pages={165724--165733},
  year={2019},
  publisher={IEEE}
}

@inproceedings{van2021multi,
  title={Multi-view analysis of unregistered medical images using cross-view transformers},
  author={Van Tulder, Gijs and Tong, Yao and Marchiori, Elena},
  booktitle={Medical Image Computing and Computer Assisted Intervention--MICCAI 2021: 24th International Conference, Strasbourg, France, September 27--October 1, 2021, Proceedings, Part III 24},
  pages={104--113},
  year={2021},
  organization={Springer}
}

@article{chen2022transformers,
  title={Transformers improve breast cancer diagnosis from unregistered multi-view mammograms},
  author={Chen, Xuxin and Zhang, Ke and Abdoli, Neman and Gilley, Patrik W and Wang, Ximin and Liu, Hong and Zheng, Bin and Qiu, Yuchen},
  journal={Diagnostics},
  volume={12},
  number={7},
  pages={1549},
  year={2022},
  publisher={MDPI}
}

@article{obeagu2024breast,
  title={Breast cancer: A review of risk factors and diagnosis},
  author={Obeagu, Emmanuel Ifeanyi and Obeagu, Getrude Uzoma},
  journal={Medicine},
  volume={103},
  number={3},
  pages={e36905},
  year={2024},
  publisher={LWW}
}

@inproceedings{zhao2020cross,
  title={Cross-view attention network for breast cancer screening from multi-view mammograms},
  author={Zhao, Xuran and Yu, Luyang and Wang, Xun},
  booktitle={IEEE International Conference on Acoustics, Speech and Signal Processing (ICASSP)},
  pages={1050--1054},
  year={2020},
  organization={IEEE}
}

@article{jouirou2019multi,
  title={Multi-view information fusion in mammograms: A comprehensive overview},
  author={Jouirou, Amira and Ba{\^a}zaoui, Abir and Barhoumi, Walid},
  journal={Information Fusion},
  volume={52},
  pages={308--321},
  year={2019},
  publisher={Elsevier}
}

@article{sun2019multi,
  title={Multi-view convolutional neural networks for mammographic image classification},
  author={Sun, Lilei and Wang, Junqian and Hu, Zhijun and Xu, Yong and Cui, Zhongwei},
  journal={IEEE Access},
  volume={7},
  pages={126273--126282},
  year={2019},
  publisher={IEEE}
}

@inproceedings{liu2021swin,
  title={Swin transformer: Hierarchical vision transformer using shifted windows},
  author={Liu, Ze and Lin, Yutong and Cao, Yue and Hu, Han and Wei, Yixuan and Zhang, Zheng and Lin, Stephen and Guo, Baining},
  booktitle={Proceedings of the IEEE/CVF international conference on computer vision},
  pages={10012--10022},
  year={2021}
}

@article{quintana2023exploiting,
  title={Exploiting patch sizes and resolutions for multi-scale deep learning in mammogram image classification},
  author={Quintana, Gonzalo I{\~n}aki and Li, Zhijin and Vancamberg, Laurence and Mougeot, Mathilde and Desolneux, Agn{\`e}s and Muller, Serge},
  journal={Bioengineering},
  volume={10},
  number={5},
  pages={534},
  year={2023},
  publisher={MDPI}
}

@article{dos2024deep,
  title={Deep Cross-Training: An Approach to Improve Deep Neural Network Classification on Mammographic Images},
  author={dos Santos, Keila Lucas and dos Santos Silva, Marcelino Pereira},
  journal={Expert Systems with Applications},
  volume={238},
  pages={122142},
  year={2024},
  publisher={Elsevier}
}

@article{yang2024mammo,
  title={Mammo-Clustering: A Weakly Supervised Multi-view Global-Local Context Clustering Network for Detection and Classification in Mammography},
  author={Yang, Shilong and Zhang, Chulong and Zang, Qi and Yu, Juan and Zeng, Liang and Luo, Xiao and Xing, Yexuan and Pan, Xin and Li, Qi and Liang, Xiaokun and others},
  journal={arXiv preprint arXiv:2409.14876},
  year={2024}
}

@inproceedings{allaoui2024hybridmammonet,
  title={HybridMammoNet: A Hybrid CNN-ViT Architecture for Multi-view Mammography Image Classification},
  author={Allaoui, Hind and Alj, Youssef and Ameskine, Yassine},
  booktitle={2024 IEEE 12th International Symposium on Signal, Image, Video and Communications (ISIVC)},
  pages={1--6},
  year={2024},
  organization={IEEE}
}

@article{liao2024open,
  title={An open codebase for enhancing transparency in deep learning-based breast cancer diagnosis utilizing CBIS-DDSM data},
  author={Liao, Ling and Aagaard, Eva M},
  journal={Scientific Reports},
  volume={14},
  number={1},
  pages={27318},
  year={2024},
  publisher={Nature Publishing Group UK London}
}

\end{document}